\documentclass[a4paper,11pt]{article}
\pdfoutput=1 
\usepackage[indent=8pt]{parskip}
\usepackage{jcappub} 

\usepackage[T1]{fontenc} 
\usepackage{enumitem}

\usepackage[english]{babel}				

\usepackage{amsmath}
\usepackage{amssymb}
\usepackage{mathtools}					
\usepackage{xcolor}						

\usepackage[colorlinks=true,urlcolor=blue,citecolor=blue,linkcolor=blue]{hyperref}

\newcommand{\Lag}{\mathcal{L}}

\newcommand\LC[1]{\mathring{#1}{}}
\newcommand\LCG{\LC{\Gamma}}
\newcommand\LCD{\LC{\nabla}}
\newcommand\LCR{\LC{R}}

\newcommand\LCten{\mathcal{E}}
\newcommand\dex{\mathrm{d}}

\newcommand\coR{\mathcal{C}}
\newcommand\homR{\mathcal{P}}

\newcommand\tT{\mathsf{t}}
\newcommand\tQ{\mathsf{q}}
\newcommand\vQ{\varLambda}

\newcommand\aT{S}
\newcommand\aR{\mathcal{H}}

\newcommand\tor[1]{\tilde{#1}{}}
\newcommand\torR{\tor{R}}
\newcommand\tord{\tor{d}}

\newcommand\WC[1]{\hat{#1}{}}
\newcommand\WCR{\WC{R}}
\newcommand\WCaR{\WC{\mathcal{H}}}
\newcommand\WCd{\WC{d}}

\newcommand\FD[1]{{}^{\scriptscriptstyle #1}\!F{}}
\newcommand\SD[1]{{}^{\scriptscriptstyle #1}\!H{}}

\title{Vector stability in quadratic metric-affine theories}

\author{Alejandro Jim\'enez-Cano}
\author{and Francisco Jos\'{e} Maldonado Torralba}

\affiliation{Laboratory of Theoretical Physics, Institute of Physics, University of Tartu, W. Ostwaldi 1, 50411 Tartu, Estonia}
\emailAdd{alejandro.jimenez.cano@ut.ee}
\emailAdd{fmaldo01@ucm.es}

\abstract{
In this work we study the stability of the four vector irreducible pieces of the torsion and the nonmetricity tensors in the general quadratic metric-affine Lagrangian in 4 dimensions. The goal will be to elucidate under which conditions the spin-1 modes associated to such vectors can propagate in a safe way, together with the graviton. This highly constrains the theory reducing the parameter space of the quadratic curvature part from 16 to 5 parameters. We also study the sub-case of Weyl-Cartan gravity, proving that the stability of the vector sector is only compatible with an Einstein-Proca theory  for the Weyl vector.

}

\begin{document}
\maketitle

\section{Introduction}

There is no doubt that the Theory of General Relativity (GR) is one of the most successful theories in Physics, due to its mathematical robustness and its numerous predictions that agree with experimental data \cite{Wald:1984rg,Will:2014kxa}. For instance, gravitational waves, which were predicted by the theory over a hundred years ago, are still being measured for the first time in our days \cite{LIGOScientific:2016aoc}. Nevertheless, some cosmological and astrophysical phenomena cannot be explained in the GR framework with just the matter content of the Standard Model of Particle Physics. For example, the Cosmic Microwave Background does not fit the GR predictions with baryonic matter \cite{Planck:2015fie}.

These problems are solved assuming GR as the correct gravitational framework and modelling the accelerated expansion with a new type of energy (\emph{dark energy}) in the form of a cosmological constant \cite{Peebles:2002gy}, and the rotation curves by adding a new form of matter, which interacts weakly and gravitationally, which is known as \emph{cold dark matter}.

Another perspective to tackle these and other problems is assuming that GR receives some modifications at the appropriate scales. We know from the Lovelock theorem that any Lorentz invariant four-dimensional local modification of the Einstein-Hilbert action of GR necessarily introduces new degrees of freedom (d.o.f.'s) at low energies \cite{Lovelock:1971yv}. Hence, we can explore if such d.o.f.'s can be used to model for instance dark matter or inflation. This line of thought has been explored in the literature for some modifications of GR (see e.g. \cite{Cembranos:2008gj,Clifton:2011jh,Amendola:2016saw,dombriz2021dark}).

When exploring such modifications we have to be careful with the fact that the new propagating d.o.f.'s may have a pathological behaviour. Consequently, we need to make sure that the modified theory is stable for sensible solutions like cosmology or Minkowski.

In the present text, we will work in the framework of Metric-Affine Gauge (MAG) gravity, which arises from a gauge procedure with respect to the affine group. The resulting spacetime manifold gets equipped with a general connection  not necessarily equal to the the Levi-Civita one. In this context, the torsion and the curvature tensors associated to the connection play the role of the gauge field strengths, although there is another natural tensor that one can construct by covariantizing the derivative of the metric, the nonmetricity tensor. It was shown for the simpler case of Poincar\'e Gauge gravity (PG), in which the nonmetricity is assumed to be zero, that the Yang-Mills-like Lagrangian does not have the appropriate Einsteinian limit \cite{Tseytlin:1981nu}.\footnote{
    Notice that the Einsteinian limit can be recovered by constructing a Yang-Mills theory in 5 dimensions (MacDowell-Mansouri formulation) \cite{MacDowell:1977jt, Wise:2006sm}.
    } 
For this reason, in PG and, by extension, in MAG, the Lagrangian is usually constructed as the most general quadratic combination in torsion, nonmetricity and curvature, and the result is known as the quadratic metric-affine Lagrangian (qMAG) \cite{Jimenez-Cano:2022arz, Jimenez-Cano:2020lea, JCthesis, Percacci:2020ddy, Marzo:2021} (sub-cases of which have been considered e.g. in \cite{Bahamonde:2021akc, Bahamonde:2021qjk, Iosifidis:2021fnq}). For a complete review of the metric-affine framework, see \cite{Hehl:1994ue} (see also \cite{JCthesis}).

We know from \cite{Percacci:2020ddy} that, apart from the usual graviton, the field content of qMAG around Minkowski space consists of several fields with spins from 0 to 2 and one spin-3 field. We clearly expect not all of them to be stable (see also the recent publication \cite{Marzo:2021}). This motivates the stability analysis of the fields involved in order to reduce the parameter space of the theory.

Some of the fields involved in qMAG are the spin-1 fields associated to the vector irreducible parts of the torsion and nonmetricity, namely the axial vector and trace of the torsion, and the two traces of the nonmetricity. The goal of this paper is to study the conditions for the stable propagation of (at least one of) {\it these four} vector fields, together with the graviton. Some of the questions we will address in this text are: Is there any sub-case of qMAG containing just the GR term and stable terms for some of these four vectors? Can the four vectors propagate safely?

In the second part of the paper we will use essentially the same techniques to analyze the Weyl-Cartan restriction of qMAG, in which the entire nonmetricity has been reduced to just one of its traces.

This article is structured as follows. In Section \ref{sec:MAGframe} we cover some basic aspects about metric-affine gravity: Section \ref{sec:MAGtools} reviews the geometrical framework of metric-affine theories, and later in Section \ref{sec:MAGlag}, we introduce the qMAG Lagrangian, which is the base of our study, as well as a convenient splitting of it. In Section \ref{sec:purevec} we study the conditions for the stable propagation of the spin-1 fields contained in the vector irreducible parts of the torsion and nonmetricity for the purely vector sector of qMAG and their implications in the tensor sector. In Section \ref{sec:Weyl} we apply our results from the general qMAG to the subclass after the evaluation in a Weyl-Cartan geometry. Finally, we collect and discuss our results in Section \ref{sec:conclusions}. At the end we include Appendix \ref{app:completevectorLag} containing the expressions of the isolated vector sectors in the general case of qMAG, and Appendix \ref{app:PGcase} in which we summarize the case of quadratic Poincar\'e Gauge gravity for completeness.

~

\noindent{\bf Notation and conventions} \\
We will use the mostly minus signature for the metric $(+,-,-,-)$ and the natural units  $c=1$. We also introduce the notation $H_{(\mu\nu)}:= \frac{1}{2!}(H_{\mu\nu}+H_{\nu\mu})$ and $H_{[\mu\nu]}:= \frac{1}{2!}(H_{\mu\nu}-H_{\nu\mu})$, and analogously for an object with $n$ indices ($2!\to n!$). For the covariant derivative, the curvature, the torsion and the nonmetricity we use the definitions \eqref{eq:nabla}, \eqref{eq:defR}, \eqref{eq:defT} and \eqref{eq:defQ}, respectively. A ring will be employed to denote objects associated to the Levi-Civita connection  ($\LCG_{\mu\nu}{}^{\rho}$, $\LCR$...), a tilde for those that correspond to a metric-compatible torsionful connection  ($\tor{\Gamma}_{\mu\nu}{}^{\rho}$, $\torR$...) and a hat for Weyl-Cartan connections   ($\WC{\Gamma}_{\mu\nu}{}^{\rho}$, $\WCR$...); by extension, we will also employ these same symbols to indicate that a certain Lagrangian has been evaluated in one of these types of connections: $\LC{\Lag}(\Gamma):=\Lag(\LC{\Gamma})$, $\tor{\Lag}(\Gamma):=\Lag(\tor{\Gamma})$ and $\WC{\Lag}(\Gamma):=\Lag(\WC{\Gamma})$.

\section{The metric-affine framework} \label{sec:MAGframe}

\subsection{Geometrical tools} \label{sec:MAGtools}

Consider a smooth manifold equipped with a linear connection $(M, \Gamma_{\mu\nu}{}^\rho)$. Such a connection can be seen as a way of comparing vectors at two arbitrary distinct points connected by a curve, which is understood as a notion of ``parallelism''. The connection defines a covariant derivative operator that acts on tensors as follows 
\begin{equation}
    \nabla_\rho H_\mu{}^\nu = \partial_\rho H_\mu{}^\nu + \Gamma_{\rho\sigma}{}^\nu H_\mu{}^\sigma - \Gamma_{\rho\mu}{}^\sigma H_\sigma{}^\nu\,.\label{eq:nabla}
\end{equation}
The associated curvature tensor is given in components by
\begin{equation}
    R_{\mu\nu\rho}{}^\lambda \coloneqq 2 \big( \partial_{[\mu}\Gamma_{\nu]\rho}{}^\lambda + \Gamma_{[\mu|\sigma}{}^\lambda \Gamma_{|\nu]\rho}{}^\sigma \big), \label{eq:defR}
\end{equation}
which has two independent contractions, respectively, the Ricci tensor and the homothetic curvature:
\begin{eqnarray}
    R_{\mu\nu} &\coloneqq &R_{\mu\lambda\nu}{}^\lambda\,, \\
    \homR_{\mu\nu} &\coloneqq& R_{\mu\nu\lambda}{}^\lambda \qquad = \frac{1}{2}(\partial_\mu Q_\nu - \partial_\nu Q_\mu) \label{eq:homothR}\,.
\end{eqnarray}
    
Another tensorial object associated to the connection is the torsion tensor,
\begin{equation}
    T_{\mu\nu}{}^\rho \coloneqq 2 \Gamma_{[\mu\nu]}{}^\rho\,,\label{eq:defT}
\end{equation}
whose trace is given by
\begin{equation}
    T_\mu \coloneqq T_{\mu\rho}{}^\rho\,.
\end{equation}

If we now add a metric to the geometry, $(M,g_{\mu\nu},\Gamma_{\mu\nu}{}^\rho)$, which will allow to measure distances and angles in the manifold, it is possible to define a new object called nonmetricity tensor,
\begin{equation}
    Q_{\rho\mu\nu} \coloneqq - \nabla_\rho g_{\mu\nu}\,,\label{eq:defQ}
\end{equation}
which has two independent traces that can be encoded in the objects:
\begin{equation}
    Q_\mu \coloneqq g^{\nu\rho}Q_{\mu\nu\rho}\,,\qquad \vQ_\mu \coloneqq g^{\nu\rho}\left(Q_{\nu\rho\mu}-\frac{1}{4}Q_\nu g_{\rho\mu} \right)\,.
\end{equation}
Furthermore, the Levi-Civita tensor of the metric  $\LCten_{\mu\nu\rho\lambda}\coloneqq  4!\delta^{[1}_\mu\delta^2_\nu\delta^3_\rho\delta^{4]}_\lambda \sqrt{|g|}$ (i.e., the canonical volume form) establishes a Hodge duality between $p$-forms and $(4-p)$-forms for all $p$'s. In particular, this associates the totally antisymmetric part of the torsion with an axial vector:
\begin{equation} 
\aT_\mu \coloneqq \frac{1}{2} T^{\nu\rho\lambda}\LCten_{\nu\rho\lambda\mu}\,.
\end{equation}
As a result of all of this, the torsion and the nonmetricity can be expressed as:
\begin{eqnarray}
  T_{\mu\nu}{}^\rho &=& \frac{2}{3}T_{[\mu}\delta_{\nu]}^\rho - \frac{1}{3}\LCten_{\mu\nu}{}^{\rho\lambda}S_\lambda + \tT_{\mu\nu}{}^\rho \,,\\
  Q_{\mu\nu\rho} &=& \frac{1}{4}Q_\mu g_{\nu\rho} + \frac{1}{9} \big[4\vQ_{(\nu} g_{\rho)\mu}-\vQ_\mu g_{\nu\rho}\big] + \tQ_{\mu\nu\rho}\,.
\end{eqnarray}
These are implicit definitions for the tracefree tensorial parts $\tT_{\mu\nu}{}^\rho$ and $ \tQ_{\mu\nu\rho}$.\footnote{Although $\tT_{\mu\nu}{}^\rho$ is an irreducible part of the torsion under the pseudo-orthogonal group, $\tQ_{\mu\nu\rho}$ admits a decomposition into two irreducible parts. However, such a splitting will not be relevant for our analysis.}

In addition, the metric also allows to construct another trace for the curvature, which we will call co-Ricci tensor,
\begin{equation}
  \coR_{\mu}{}^\nu \coloneqq g^{\rho\lambda} R_{\mu\rho\lambda}{}^\nu \,,
\end{equation}
as well as a one scalar and one pseudo-scalar quantities
\begin{equation}
    R \coloneqq g^{\mu\nu}R_{\mu\nu}\,,\qquad \aR \coloneqq \LCten_{\mu\nu\rho\lambda}R^{\mu\nu\rho\lambda}\,.
\end{equation}

Finally, it is worth recalling that any metric induces naturally a connection over the manifold, called the Levi-Civita connection (for it and its associated objects we use the notation $\LCG_{\mu\nu}{}^\rho$, $\LCR_{\mu\nu\rho}{}^\lambda$ and $\LCD_\mu$). This connection is uniquely determined by the torsion-free ($T_{\mu\nu}{}^\rho=0$) and the metric-compatibility ($Q_{\mu\nu\rho}=0$) conditions, as can be immediately deduced from the identity
\begin{equation}
    \Gamma_{\mu\nu}{}^{\rho}-\LCG_{\mu\nu}{}^{\rho}=\Xi_{\mu\nu}{}^{\rho},\qquad \text{where}\qquad 2 g_{\sigma\rho}\Xi_{\mu\nu}{}^\sigma  \coloneqq T_{\mu\nu\rho} + T_{\rho\mu\nu} - T_{\nu\rho\mu} + Q_{\mu\nu\rho}  + Q_{\nu\rho\mu} - Q_{\rho\mu\nu}\,.
\end{equation}
Here $\Xi_{\mu\nu}{}^\rho$ is called distorsion tensor and allows to expand the curvature of the general connection in terms of the Levi-Civita Riemann tensor:
\begin{equation}
    R_{\mu\nu\rho}{}^\lambda = \LCR_{\mu\nu\rho}{}^\lambda + 2\,\LCD_{[\mu}\Xi_{\nu]\rho}{}^\lambda + 2\, \Xi_{[\mu|\sigma}{}^\lambda \Xi_{|\nu]\rho}{}^\sigma\,.\label{eq:postRdec}
\end{equation}

\subsection{Quadratic metric-affine Lagrangian} \label{sec:MAGlag}

Consider now a metric-affine theory of the type
\begin{equation}
    S_\text{MAG}[g_{\mu\nu},\Gamma_{\mu\nu}{}^\rho]= \int \Lag_\text{MAG}(g_{\mu\nu}, T_{\mu\nu}{}^\rho, Q_{\mu\nu\rho},R_{\mu\nu\rho}{}^\lambda) \ \sqrt{|g|}\, \dex^4x\,.
\end{equation}
For the gravitational Lagrangian we choose the most general one fulfilling the following requirements: (1) it only contains algebraic invariants depending on the curvature, the torsion and the nonmetricity tensor (in addition to the cosmological constant term); (2) its terms are at most quadratic in these quantities; (3) no odd-parity invariants are considered. The resulting Lagrangian can be parameterized as
\begin{eqnarray}
    2\kappa \Lag_\text{MAG} &\coloneqq& -2\kappa\Lambda + a_0 R + a_1 \tT_{\mu\nu\rho} \tT^{\mu\nu\rho} + a_2 T_\mu T^\mu + a_3 \aT_\mu \aT^\mu + b_1 \tQ_{\mu\nu\rho} \tQ^{\mu\nu\rho} + b_2 \tQ_{\mu\nu\rho} \tQ^{\rho\mu\nu}
    \nonumber \\
&&  + b_3 Q_\mu Q^\mu+ b_4 \vQ_\mu \vQ^\mu + b_5 Q_\mu \vQ^\mu + c_1 \tQ_{\mu\nu\rho} \tT^{\mu\nu\rho} + c_2 Q_\mu T^\mu +  c_3 \vQ_\mu T^\mu\nonumber \\
&&  +\ell^2 \Big[ R^{\mu\nu\rho\lambda} \big(d_1R_{\mu\nu(\rho\lambda)}+ d_2R_{\mu\nu[\rho\lambda]} + d_3 R_{\mu(\rho\lambda)\nu}+ d_4 R_{\mu[\rho\lambda]\nu}+ d_5 R_{\rho\lambda\mu\nu}\big)\nonumber\\
&&\qquad + R^{\mu\nu} (d_6R_{(\mu\nu)}+d_7R_{[\mu\nu]})+\coR^{\mu\nu}(d_8 R_{(\mu\nu)} +d_9 R_{[\mu\nu]})+\coR^{\mu\nu}(d_{10} \coR_{(\mu\nu)} +d_{11} \coR_{[\mu\nu]}) \nonumber\\
&&\qquad + \homR^{\mu\nu} (d_{12}R_{\mu\nu}+d_{13}\coR_{\mu\nu}+d_{14}\homR_{\mu\nu}) +d_{15} R^2+ d_{16}\aR^2\Big]\,,
\label{eq:LagMAG}
\end{eqnarray}
where $\kappa\coloneqq 8\pi G$ is the Einstein constant, $\Lambda$ is the cosmological constant, $\ell$ is a parameter with dimensions $[\ell]=\mathsf{L}$, and the rest of the parameters ($a_0,a_1,...,a_3,b_1,...,b_5,c_1,...,c_3, d_1,...,d_{16}$) are dimensionless.  

After performing the post-Riemannian decomposition \eqref{eq:postRdec}, the general Lagrangian \eqref{eq:LagMAG} can be split as follows:
\begin{equation}
    \Lag_\text{MAG} = \Lag_\text{GR} + \Lag_{\LCR\LCR} + \Lag_\text{v} + \Lag_*\,, \label{eq:LagSplit}
\end{equation}
where
\begin{align}
    \Lag_\text{GR}&\coloneqq-\Lambda+\frac{a_0}{2\kappa}\LCR\,,\\
    \Lag_{\LCR\LCR}&\coloneqq \frac{\ell^2}{2\kappa}\left[d_{15}\LCR^2+(d_6-d_8+d_{10})\LCR_{\mu\nu}\LCR^{\mu\nu} + \left(d_2-\frac{d_4}{2}+d_5\right)\LCR_{\mu\nu\rho\lambda}\LCR^{\mu\nu\rho\lambda}\right]\,,
\end{align}
$\Lag_\text{v}$ is the part that depends exclusively on the vector variables $\{T_\mu, S_\mu, Q_\mu, \vQ_\mu\}$ and $\Lag_*$ is the remaining piece, which contains the terms involving only the tensor variables  $\{\tQ_{\mu\nu\rho}, \tT_{\mu\nu}{}^\rho\}$ and their couplings to the vectors. Moreover, it will be also convenient to further expand the pure vector sector as follows:
\begin{equation}
    \Lag_\text{v} = \Lag_T + \Lag_S + \Lag_Q+ \Lag_\vQ + \Lag_\text{v, mix}
\end{equation}
with
\begin{eqnarray}
     \Lag_T  &\coloneqq&\Lag_\text{v}|_{S_\mu=Q_\mu=\vQ_\mu=0}\,,\nonumber\\
     \Lag_S  &\coloneqq&\Lag_\text{v}|_{T_\mu=Q_\mu=\vQ_\mu=0}\,,\nonumber\\
     \Lag_Q  &\coloneqq&\Lag_\text{v}|_{S_\mu=T_\mu=\vQ_\mu=0}\,,\nonumber\\
     \Lag_\vQ&\coloneqq&\Lag_\text{v}|_{S_\mu=T_\mu=Q_\mu=0}\,, \label{eq:restrictVecL}
\end{eqnarray}
and $\Lag_\text{v, mix}$ being the sector containing interactions between the different vectors.

For future purposes we also introduce the metric-affine Gauss-Bonnet invariant 
\begin{equation}
   \Lag_\text{GB}\ :=\ 3! R_{[\mu\nu}{}^{\mu\nu} R_{\rho\lambda]}{}^{\rho\lambda}  \ =\ R^2-(R_{\mu\nu}-\coR_{\mu\nu})(R^{\nu\mu}-\coR^{\nu\mu})+R_{\mu\nu\rho\lambda}R^{\rho\lambda\mu\nu}\,,
\end{equation}
which is guaranteed to be a boundary term if $\vQ_\mu=0=\tQ_{\mu\nu\rho}$ \cite{JJGB} (in particular, in Poincar\'e and Weyl-Cartan gravity).

\section{Vector sector of quadratic MAG with propagating spin-1 fields} \label{sec:purevec}

In this section, which is the central one of this article, we are going to perform a general analysis of the vector sector $\Lag_\text{v}$. We recall one more time that the goal of the present text is to find sub-cases of \eqref{eq:LagMAG} that propagate the usual graviton together with at least one of the spin-1 fields in $\{T_\mu, S_\mu, Q_\mu,\vQ_\mu\}$, in a safe way. The idea would be to find necessary conditions for such a theory by just studying the vector sector. Our approach has been inspired by the one performed in \cite{BeltranMaldonado2020}, and based on the stability conditions required for vectors  \cite{BeltranJimenez:2016rff,Heisenberg:2014rta,GenMultiProca,Heisenberg:2016eld}. Now we enumerate the steps we will follow:
\begin{enumerate}
    \item Restrict the parameters so that the Einstein-Hilbert (plus Gauss-Bonnet) Lagrangian is recovered when torsion and nonmetricity are vanishing.
    \item Study the stability of each of the vector fields separately.
    \item Ensure healthy interactions between vectors.
    \item Constrain appropriately the kinetic matrix to avoid ghosts.
\end{enumerate}
The conditions coming from the steps 1., 2. and 3. are necessary in order to have generically safe vectors. Regarding the last step, it is true that the analysis of the kinetic matrix would require to study not only the vector sector, but also its kinetic couplings to the tensor pieces contained in $\Lag_*$. However, as we will explain below, the conditions that we will obtain from the analysis of this {\it partial} kinetic matrix continue being necessary conditions in the full theory.

The first step will be described in Section \ref{sec:RiemmLimit}. In Section \ref{sec:convenientbasis}, we will introduce some useful notation and expressions to eliminate certain invariants from the action. Then, in Sections \ref{sec:MAGselfstab}, \ref{sec:MAGmixstab} and \ref{sec:kinMAG}, we develop the second, third and fourth steps of our stability analysis, respectively. Finally, in Section \ref{sec:implicfullMAG}, we see the implications of this analysis in the full theory, including $\Lag_*$.

\subsection{Constraints on the parameters from the Riemannian limit}\label{sec:RiemmLimit}

When setting the torsion and nonmetricity to zero in \eqref{eq:LagSplit}, only the first two terms survive, and these can be rearranged as
\begin{equation}
    2\kappa \big(\Lag_\text{GR} + \Lag_{\LCR\LCR}\big) = -2\kappa\Lambda + a_0 \LCR +\ell^2\left[\beta\mathring{R}^2+\alpha\mathring{R}_{\mu\nu}\mathring{R}^{\mu\nu} \right]+\ell^2 \left(d_2-\frac{d_4}{2}+d_5\right)\LC{\Lag}_\text{GB}.
\label{eq:MAGLClimit}
\end{equation}
Here we have introduced  the Levi-Civita Gauss-Bonnet invariant, $\LC{\Lag}_\text{GB}:=\Lag_\text{GB}|_{\Gamma = \LCG}$, which can be dropped since we are working in four dimensions, and the parameters
\begin{eqnarray}
    \beta&\coloneqq&d_{15}-d_2+\frac{d_4}{2}-d_5\,,\nonumber\\
    \alpha&\coloneqq&d_{10}+4d_2-2d_4+4d_5+d_6-d_8\,.\label{eq:defalbe}
\end{eqnarray}
From now on we will work in terms of $\alpha$ and $\beta$ instead of $d_2$ and $d_6$.

In order to avoid extra potentially pathological d.o.f.'s coming from higher curvature terms, we shall require recovering just the Einstein-Hilbert Lagrangian when torsion and nonmetricity are zero. Consequently, we have to impose\footnote{In \cite{BeltranMaldonado2020}, it was shown that relaxing these requirements to just $\alpha=0$, allows for stable Poincar\'e theories. Such theories do not propagate spin-1 fields but an additional scalar $\chi$ coming from the longitudinal part of the torsion trace ($T_\mu = \partial_\mu \chi$).}
\begin{equation}
    \alpha = 0\,,\qquad \beta=0 \,.\label{cond:1}
\end{equation}
These conditions are equivalent to the eliminations,
\begin{equation}
    \text{(I)}\coloneqq\begin{cases}
d_6 \to -d_{10}-4d_2+2d_4-4d_5+d_8,\\
d_2 \to d_{15}-d_5+\frac{d_4}{2}.\\
\end{cases}\label{cond:3}
\end{equation}

\subsection{Convenient basis of operators}\label{sec:convenientbasis}

In order to have more compact expressions, we will define the symmetric and antisymmetric derivatives of a covector $V_{\mu}$ with respect to the Levi-Civita connection as
\begin{equation}
    \SD{V}_{\mu\nu}\coloneqq2\mathring{\nabla}_{(\mu}V_{\nu)},\qquad \FD{V}_{\mu\nu}\coloneqq2\mathring{\nabla}_{[\mu}V_{\nu]}\quad (\equiv 2\partial_{[\mu}V_{\nu]}).
\end{equation}
In addition, we will use $\SD{V}$ for the trace of $\SD{V}_{\mu\nu}$, i.e., for the divergence of the vectors: $\SD{V}=2\mathring{\nabla}_{\mu}V^{\mu}$.
This kind of notation makes it easier to express the safe interactions of a single vector field, which are given in \cite{BeltranJimenez:2016rff}. 

Let $V_\mu$ and $W_\mu$ be two arbitrary but different elements of $\{T_\mu,S_\mu,Q_\mu,\vQ_\mu\}$. It can be shown that the terms with field and derivative content $VVV\LCD$ are connected via a boundary term and, similarly, those of the type $VVW\LCD$ can be split into two families of terms related via boundary terms. The explicit expressions for these three boundary terms are:
\begin{eqnarray}
    2\LCD_\mu(V^\mu V_\nu V^\nu)&=& \SD{V} V_\nu V^\nu + 2\SD{V}_{\mu\nu}  V^\mu V^\nu \,,\\
    2\LCD_\mu(W^\mu V_\nu V^\nu)&=& \SD{W} V_\nu V^\nu +2 (\SD{V}_{\mu\nu}+\FD{V}_{\mu\nu})  W^\mu V^\nu \,,\\
    2\LCD_\mu(V^\mu W_\nu V^\nu)&=& \SD{V} W_\nu V^\nu + \SD{W}_{\mu\nu} V^\mu V^\nu + (\SD{V}_{\mu\nu}-\FD{V}_{\mu\nu})  W^\mu V^\nu \,,
\end{eqnarray}
which allow to perform the elimination of the terms containing the symmetric parts in favour of the divergences $\SD{V}$ and the antisymmetric parts $\FD{V}_{\mu\nu}$, so that we just keep the invariants of the types:
\begin{equation}
    \big\{ \SD{V} V_\nu V^\nu,\ \SD{W} V_\nu V^\nu,\ \SD{V} W_\nu V^\nu,\  \FD{V}_{\mu\nu}  W^\mu V^\nu  \big\}.
\end{equation}

Regarding the terms with content $VW\LCD\LCD$, the boundary term
\begin{equation}
   2\LCD_\mu \big( V^{[\mu}\LCD_\nu W^{\nu]}\big)=
    \frac{1}{2}\LCR\,V_{\mu}W^{\mu}+\frac{1}{4}\left(\SD{V}\SD{W}-\SD{V}_{\mu\nu}\SD{W}^{\mu\nu}\right)+\LC{G}_{\mu\nu}V^{\mu}W^{\nu}+\frac{1}{4}\FD{V}_{\mu\nu}\FD{W}^{\mu\nu}\label{eq:boundterm4}
\end{equation}
can be used to remove the terms $\SD{V}_{\mu\nu}\SD{W}^{\mu\nu}$.

\subsection{Independent healthiness of the spin-1 fields}\label{sec:MAGselfstab}

We assume that the spin-1 part of the vectors need to propagate safely when considered independently, namely, that each of $\Lag_T$, $\Lag_S$, $\Lag_Q$ and $\Lag_\vQ$  (defined in \eqref{eq:restrictVecL}) must be a healthy Lagrangian for a spin-1 field. Their explicit expressions (up to boundary terms) can be found in Appendix \ref{app:completevectorLag}. From now on, due to the reasons explained in Section \ref{sec:RiemmLimit}, we impose $\alpha=\beta=0$.

For the case of the MAG Lagrangian, the possible stable terms for any vector $V^{\mu}$ are of the following form \cite{BeltranJimenez:2016rff} (see also \cite{Heisenberg:2014rta,GenMultiProca,Heisenberg:2016eld})
\begin{eqnarray}
\Omega_1&=&\alpha_1f\left(V_{\mu}V^{\mu}\right)+\alpha_2\FD{V}_{\mu\nu}\FD{V}^{\mu\nu}+\alpha_3\LC{G}_{\mu\nu}V^{\mu}V^{\nu},\\
\Omega_2&=&\alpha_4\SD{V}V_{\mu}V^{\mu}+\alpha_5\SD{V}_{\mu\nu}V^{\mu}V^{\nu},\\
\Omega_3&=&\LCR\,V_{\mu}V^{\mu}+\dfrac{1}{2}\left(\SD{V}^2-\SD{V}_{\mu\nu}\SD{V}^{\mu\nu}\right),\label{eq:stab3}
\end{eqnarray}
where the $\alpha_i$'s are arbitrary coefficients. Regarding non-minimal couplings to curvature, observe that only the combination $\LC{G}_{\mu\nu}V^{\mu}V^{\nu}$ is safe. The coupling to the Ricci scalar in \eqref{eq:stab3} is pathological \cite{Himmetoglu:2008zp, BeltranJimenez:2008zzi, Armendariz-Picon:2009kfd, Himmetoglu:2009qi}, as well as the other two terms, $\SD{V}^2$ and $\SD{V}_{\mu\nu}\SD{V}^{\mu\nu}$, except when they appear in such a specific combination. Indeed, $\Omega_3$ is equal to a combination of safe terms plus a total derivative due to the equality \eqref{eq:boundterm4} with $V^\mu=W^\mu$.

Now we proceed with the study of the stability of the vectors separately:

\begin{enumerate}

\item \textit{Trace torsion vector}. In this case the restriction of the Lagrangian after imposing the substitutions (I) we introduced in \eqref{cond:3}, has a Proca form, 
\begin{equation}
    2\kappa\Lag_T|_\text{(I)}
    =\frac{1}{9}\ell^2(4d_5-d_4+d_7-d_9+d_{11}-4d_{15})\FD{T}_{\mu\nu}\FD{T}^{\mu\nu}
      +\left(\frac{2a_0}{3}+a_2\right)T_{\mu}T^{\mu},
\label{lag:Ttr:I}
\end{equation}
and it does not contain any dangerous couplings. Of course the signs of the coefficients should be appropriate to avoid ghosts and/or tachyons, but in principle there is enough freedom in the parameters to do that.

\item \textit{Weyl trace vector}. Here the situation is pretty similar to the previous one. The resulting Lagrangian is
\begin{equation}
    2\kappa\Lag_Q|_\text{(I)}
    =\frac{1}{16}\ell^2(d_1+d_5+d_7+2d_{12}+4d_{14}-d_{15})\FD{Q}_{\mu\nu}\FD{Q}^{\mu\nu}+\left(\frac{3a_0}{32}+b_3\right)Q_{\mu}Q^{\mu}.
\label{lag:Qtr1:I}
\end{equation}
Notice that the stability of these two traces is in agreement with the findings in \cite{BeltranJimenez:2015pnp,BeltranJimenez:2016wxw} for the 4-dimensional case.

\item \textit{Axial torsion vector}. The restriction of the Lagrangian in this case is given by
\begin{eqnarray}
    2\kappa\Lag_S|_\text{(I)}
    &=&\frac{1}{36}\ell^2(-8d_5+2d_4-d_7+d_9-d_{11}+4d_{15})\FD{S}_{\mu\nu}\FD{S}^{\mu\nu}
      +\left(a_3-\frac{a_0}{6}\right)S_{\mu}S^{\mu}\nonumber \\
    &&-\frac{1}{9} \ell^2(4d_5-d_4)\LC{G}_{\mu\nu}S^{\mu}S^{\nu}
      +\frac{1}{24}\ell^2(24d_{16}-2d_5-d_4)\SD{S}^2\nonumber \\
    &&
      -\frac{1}{18} \ell^2(4d_5-d_4) \LCR\, S_\mu S^\mu.
\label{lag:Taxial:I}
\end{eqnarray}
The only remaining unhealthy terms are the ones related to the symmetric part of the derivative and the non-minimal coupling to the Ricci scalar. Therefore, to ensure stability we must impose
\begin{equation}
    \begin{cases}
0=4d_5-d_4,\\
0=d_5-4d_{16}.\\
\end{cases}\label{cond:axial}
\end{equation}

\item \textit{Second trace of the nonmetricity}. In this case, after restricting the Lagrangian, we obtain
\begin{eqnarray}
    2\kappa\Lag_\vQ|_\text{(I)}
    &=&\frac{1}{81}\ell^2(7d_1+9d_5-d_3+2d_8+12d_{10}+9d_{11}-d_{15})\FD{\vQ}_{\mu\nu}\FD{\vQ}^{\mu\nu}
      \nonumber \\
    &&+\left(b_4-\frac{a_0}{18}\right)\vQ_{\mu}\vQ^{\mu}+\frac{1}{2187}\ell^2(36d_1+18d_3+33d_8+78d_{10}-8d_{15}) (\vQ_{\mu}\vQ^{\mu})^2\nonumber \\
    &&+\frac{1}{54}\ell^2(2d_1+d_3+ d_8+ 6d_{10}+4d_{15})\SD{\vQ}^2\nonumber \\
    &&+\frac{2}{81} \ell^2(4d_1+2d_3+5d_8+6d_{10}-6d_{15})\LCR\, \vQ_\mu \vQ^\mu \nonumber \\
    &&+\frac{1}{9}\ell^2(d_8-2d_{10}-4d_{15})\SD{\vQ}\LCR\,\nonumber \\
    &&+\frac{8}{81}\ell^2(2d_1+d_3+4d_8-7d_{15})\LC{G}_{\mu\nu}\vQ^\mu\vQ^\nu  \nonumber \\
    &&-\frac{1}{486}\ell^2(24d_1+12d_3+17d_8+62d_{10}+4d_{15})\SD{\vQ}\vQ_\mu \vQ^\mu .
\label{lag:Qtr2:I}
\end{eqnarray}
Here, the vanishing of the coefficients of the unhealthy terms $\SD{\vQ}^2$, $\SD{\vQ}\LCR$ and $\LCR\, \vQ_\mu \vQ^\mu$, lead to the following set of conditions:
\begin{equation}
    \begin{cases}
0=2d_1+d_3+8d_{10},\\
0=d_8-2d_{10},\\
0=d_{15},
\end{cases}\label{cond:tr2Q}
\end{equation}
\end{enumerate}

The stability conditions that we have imposed so far, i.e., (I) (see \eqref{cond:3}) together with \eqref{cond:axial} and \eqref{cond:tr2Q}, lead us to the following parameter eliminations in the original Lagrangian:
\begin{eqnarray}
 \text{(II)}&\coloneqq& \Big\{ d_2\to d_5,\ d_3\to -2(d_1+4d_{10}),\ d_4\to 4d_5,\ d_6\to d_{10},\nonumber \\
 & &\quad d_8\to 2d_{10},\ d_{15}\to 0,\ d_{16}\to \frac{1}{4} d_5\Big\}.
    \label{condII}
\end{eqnarray}

\subsection{Additional stability conditions from the mixed sector}\label{sec:MAGmixstab}

After the previous analysis, only the parameters $\{ d_1,d_5,d_7,d_9,d_{10},d_{11},d_{12},d_{13},d_{14}\}$ survive for the curvature-square sector. Under these restrictions the isolated vector sectors \eqref{lag:Taxial:I},\eqref{lag:Ttr:I}, \eqref{lag:Qtr1:I} and \eqref{lag:Qtr2:I} become Proca Lagrangians:
\begin{eqnarray}
    2\kappa\Lag_S|_\text{(II)}
    &=&-\frac{1}{36}\ell^2(d_7-d_9+d_{11})\FD{S}_{\mu\nu}\FD{S}^{\mu\nu}  +\left(a_3-\frac{a_0}{6}\right)S_{\mu}S^{\mu}\,,\label{lag:Taxial:II}\\
    2\kappa\Lag_T|_\text{(II)}
    &=&\,\,\frac{1}{9}\ell^2(d_7-d_9+d_{11})\FD{T}_{\mu\nu}\FD{T}^{\mu\nu}     +\left(\frac{2a_0}{3}+a_2\right)T_{\mu}T^{\mu}\,,\label{lag:Ttr:II}\\
    2\kappa\Lag_ Q|_\text{(II)}
    &=&\frac{1}{16}\ell^2(d_1+d_5+d_7+2d_{12}+4d_{14})\FD{Q}_{\mu\nu}\FD{Q}^{\mu\nu}     +\left(\frac{3a_0}{32}+b_3\right)Q_{\mu}Q^{\mu}\,,\label{lag:Qtr1:II}\\
    2\kappa\Lag_\vQ|_\text{(II)}
    &=&\frac{1}{27}\ell^2(3d_1+3d_5+8d_{10}+3d_{11})\FD{\vQ}_{\mu\nu}\FD{\vQ}^{\mu\nu}     +\left(b_4-\frac{a_0}{18}\right)\vQ_{\mu}\vQ^{\mu}\,.\label{lag:Qtr2:II}
\end{eqnarray}

If we define
\begin{equation}
    \gamma\coloneqq d_1+2d_5+4d_{10},
\end{equation}
and use it to eliminate $d_{10}$, the mixed sector can be expressed:
\begin{eqnarray}
    &&\!\!\!\! 2\kappa\Lag_\text{v,mix}|_\text{(II)}=\nonumber\\
    &&\  \frac{1}{12}\ell^2(2d_5+d_9+2d_{13}-2\gamma)\FD{\vQ}_{\mu\nu}\FD{Q}^{\mu\nu}+\frac{1}{9}\ell^2(\gamma-d_9+2d_{11})\FD{\vQ}_{\mu\nu}\FD{T}^{\mu\nu} \nonumber\\
    &&\  -\frac{1}{12}\ell^2(\gamma+2d_7-d_9+2d_{12}-2d_{13})\FD{Q}_{\mu\nu}\FD{T}^{\mu\nu}+\frac{2}{81}\ell^2(\gamma-3d_7+3d_{11}) \FD{S}_{\mu\nu}\vQ^\mu S^\nu\nonumber\\
    &&\  +\frac{1}{36}\ell^2 (4d_5+2d_7+d_9+2d_{12}+2d_{13}-3\gamma) (\star\FD{Q})_{\mu\nu}\vQ^\mu S^\nu\nonumber\\
    &&\ +\frac{1}{81}\ell^2 (4d_1-4d_5+3d_9+6d_{11}+5\gamma) (\star\FD{\vQ})_{\mu\nu}\vQ^\mu S^\nu+\frac{2}{27}\ell^2 (\gamma-d_7+d_{11}) (\star\FD{T})_{\mu\nu}\vQ^\mu S^\nu\nonumber\\
    &&\ +\left(b_5-\frac{a_0}{4}\right)Q_\mu \vQ^\mu + \left(c_2-\frac{a_0}{2}\right)Q_\mu T^\mu +\left(c_3+\frac{2a_0}{3}\right)\vQ_\mu T^\mu \nonumber\\
    &&\  +\frac{1}{81}\ell^2\gamma\left[
    \frac{8}{3}\vQ_\mu T^\mu  S_\nu S^\nu +\frac{4}{3} S_\mu T^\mu  S_\nu\vQ^\nu -\frac{1}{2} S_\mu Q^\mu  S_\nu \vQ^\nu - \vQ_\mu Q^\mu S_\nu S^\nu -\SD{S}\vQ_\mu S^\mu -2\SD{\vQ} S_\mu S^\mu \right]\nonumber\\
    &&\ +\frac{2}{729}\ell^2\big(10d_1-8d_5+9(d_7+d_9+d_{11})+7\gamma\big)(\vQ_\mu S^\mu)^2 \nonumber\\
    &&\  -\frac{2}{729}\ell^2\big(16d_1-20d_5+9(d_7+d_9+d_{11})+4\gamma\big)\vQ_\mu\vQ^\mu S_\nu S^\nu
\end{eqnarray}
where $(\star \FD{V})^{\mu\nu}\coloneqq \frac{1}{2} \FD{V}_{\alpha\beta} \LCten^{\alpha\beta \mu\nu}$.

In order to avoid the propagation of ghostly modes, we need to impose $\gamma =0$, since this is the only way to remove the dangerous combination $\SD{S}\vQ_\mu S^\mu +2\SD{\vQ} S_\mu S^\mu$. To better understand why such a combination is problematic, consider the object
\begin{equation}
    \mathcal{J}:=\SD{V}W_\mu V^\mu + \chi~ \SD{W} V_\mu V^\mu\,,
\end{equation}
where $V_\mu$ and $W_\mu$ are generic (but different) vector fields and $\chi$ is some real parameter. After introducing St\"uckelberg fields and taking an appropriate decoupling limit, we will effectively have $V_\mu\to\partial_\mu v$ and $W_\mu\to\partial_\mu w$, and therefore we get (in Minkowski spacetime):
\begin{eqnarray}
  \mathcal{J} \to&& \square v \partial_\mu w \partial^\mu v + \chi~ \square w \partial_\mu v \partial^\mu v  \nonumber\\
  &=& \ddot{v} \dot{v} \dot{w} + \chi \ddot{w} \dot{v}^2 +\ldots \nonumber\\
  &=&    (1-2\chi) \ddot{v} \dot{v} \dot{w} + \ldots,
\end{eqnarray}
where in the last step we have integrated by parts in the time direction. We see that $\mathcal{J}$ gives a non-trivial contribution to the Hessian matrix and, consequently, allows for an Ostrogradski mode in flat spacetime unless $\chi= 1/2$. In our case, we have the combination $\SD{S}\vQ_\mu S^\mu +2\SD{\vQ} S_\mu S^\mu$, which means $\chi=2$. Hence, we need to impose $\gamma=0$ to ensure stability. Together with the previous substitution rules, we get
\begin{align}
    \text{(III)}&\coloneqq\text{(II)}\cup\Big\{ d_{10} \to -\frac{1}{4}(d_1+2d_5) \Big\}\nonumber\\
     &=\Big\{d_2\to d_5,\ d_3\to 4d_5,\ d_4\to 4d_5,\  d_6\to -\frac{1}{4}(d_1+2d_5),\nonumber\\
     &\qquad d_8\to -\frac{1}{2}(d_1+2d_5),\ d_{10} \to -\frac{1}{4}(d_1+2d_5),\ d_{15}\to 0,\ d_{16}\to \frac{1}{4} d_5\Big\}
     \label{eq:condIII}\,.
\end{align}
It is interesting to observe that the pathological terms that we have dropped in Sections \ref{sec:MAGselfstab} and \ref{sec:MAGmixstab} involve either the axial vector $S_\mu$, the second trace of the nonmetricity $\vQ_\mu$, or both, but neither the trace torsion $T_\mu$ nor the first trace of the nonmetricity $Q_\mu$.

\subsection{Analysis of the kinetic matrix for purely vector MAG}\label{sec:kinMAG}

The vector sector of the Lagrangian has the form
\begin{equation}
    \Lag_\text{v}|_\text{(III)}= \Lag_\text{FF}+\Lag_\text{FVV}+\Lag_\text{VV}+\Lag_\text{VVVV}\,, \label{eq:LagIII}
\end{equation}
where the subscripts indicate the number of field strengths and vectors in each term. We omit the explicit expressions after imposing (III), which can be immediately obtained from the ones above.

Let us ignore for a moment the tensor sector $\Lag_*$. In Section \ref{sec:implicfullMAG}, we will see that our conclusions remain valid even when including it. Having said this, the only remaining analysis is to study the kinetic matrix for the vector propagation. We need to check that the eigenvalues of the matrix are of the correct sign, so that none of them is a ghost. If we make the following identification
\begin{align}
   K_1&\coloneqq \frac{1}{6}(2d_7-d_9+2d_{12}-2d_{13}),
   &K_4&\coloneqq -\frac{1}{6}(2d_5+d_9+2d_{13}),\nonumber\\
   K_2&\coloneqq \frac{2}{9}(d_9-2d_{11}),
   &K_5&\coloneqq -\frac{4}{27}(d_1-d_5+3d_{11}),\nonumber\\
   K_3&\coloneqq -\frac{1}{4}(d_1+d_5+d_7+2d_{12}+4d_{14}),
   &\zeta&\coloneqq \frac{1}{9}(d_7-d_9+d_{11}),
\end{align}
we can then write the kinetic sector as
\begin{equation}
    \Lag_\text{FF} = -\frac{1}{4}\ell^2\ (\mathcal{K})_{ij}\FD{(i)}_{\mu\nu}\FD{(j)}^{\mu\nu},\qquad \FD{(i)}\coloneqq(\FD{S},\FD{T},\FD{Q},\FD{\vQ})\label{eq:LagFF}
\end{equation}
where we have introduced the kinetic matrix
\begin{equation}
    \mathcal{K}\coloneqq\left(\begin{array}{cccc}
\zeta & 0 & 0 & 0\\
0 & -4\zeta & K_{1} & K_{2}\\
0 & K_{1} & K_{3} & K_{4}\\
0 & K_{2} & K_{4} & K_{5}
\end{array}\right).\label{eq:KinMatrix}
\end{equation}
Note that the absence of kinetic coupling between the axial vector and the rest is justified by the fact that the Lagrangian is parity preserving.

In order to have a ghost-free theory there cannot be negative eigenvalues for this matrix, so that, after diagonalization and normalization of the fields, we get a $-1/4$ in front of each $F^2$ term in \eqref{eq:LagFF}. Of course, some of the fields can be non-propagating and, in such a case, the kinetic matrix will be positive semi-definite. At this point, it is convenient to recall some elemental criteria for positive definite and positive semi-definite matrices:
\begin{itemize}
    \item A matrix is positive definite (semi-definite) if and only if all of its principal minors are positive (respectively non-negative). In particular, the elements of the diagonal must be positive (respectively non-negative).\footnote{A principal minor is a minor calculated after eliminating some columns ($i$-th, $j$-th,...) and the rows with the same label ($i$-th, $j$-th,...). In particular, the elements of the diagonal are principal minors. }
\end{itemize}
However, for positive definite matrices there is a stronger result:
\begin{itemize}
    \item Sylvester's criterion: a matrix is positive definite if and only if all of its leading (starting in the upper-left corner) minors are positive.
\end{itemize}
We will use this in the following subsections to detect ghostly behaviors in the kinetic matrix.

\subsubsection{Implications of a propagating axial torsion}\label{sec:implicaxial}

The fact that the kinetic term of the axial torsion is decoupled from the rest allows to extract some immediate conclusions about the implications of its presence in the spectrum of the theory. Consider that the axial vector propagates in a healthy way. Since it is decoupled from the rest, this is true if and only if $\zeta>0$. Let us now focus on the kinetic matrix that mixes the rest of the vectors:
\begin{equation}
\mathcal{K}_{(3)}=\left(\begin{array}{ccc}
-4\zeta (<0) & K_{1} & K_{2}\\
K_{1} & K_{3} & K_{4}\\
K_{2} & K_{4} & K_{5}
\end{array}\right).
\end{equation}
Here there are two cases depending on how many of these three vectors propagate:
\begin{enumerate}
    \item {\it Some of them propagate}. Since the first element of the diagonal fails to be non-negative, at least one of the vectors is a ghost. 
    \item {\it None of them propagate}. This is only possible if $\mathcal{K}_{(3)}=0$, which contradicts the hypothesis $\zeta>0$.
\end{enumerate}
Notice that the latter implies that the axial torsion cannot propagate alone.

This analysis leads us to a very strong restriction: {\it a ghost-free setting with propagating axial torsion $S^\mu$ cannot be realized by any set of parameters of the theory} \eqref{eq:LagIII}. Therefore, it is not possible to have a subclass of  \eqref{eq:LagIII} with four healthy propagating vectors.

\subsubsection{Purely vector MAG with three propagating spin-1 fields}\label{sec:3vectors}

The conclusions of Section \ref{sec:implicaxial} enforce the elimination of the kinetic term of the axial torsion vector by imposing 
\begin{equation}
    \zeta =0 \qquad (\Leftrightarrow\quad d_7-d_9+d_{11}= 0).
\end{equation}
Accordingly, now the kinetic matrix reads
\begin{equation}
\left(\begin{array}{ccc}
0 & K_{1} & K_{2}\\
K_{1} & K_{3} & K_{4}\\
K_{2} & K_{4} & K_{5}
\end{array}\right),\label{eq:K3zeta0}
\end{equation}
where the coefficients $\{K_i\}$ continue being independent. Notice that this matrix cannot be positive definite (neither negative definite) because the first element of the diagonal is not a positive number. If we assume that the three remaining vectors propagate (i.e. that none of the eigenvalues is vanishing), then the matrix must be indefinite and, therefore, there is at least one ghost.

We then conclude that {\it the situation in which three vectors are propagating (none of them being a ghost) cannot be realized by any set of parameters of the theory} \eqref{eq:LagIII}. 

Observe that the situation is analogous to Ricci-Based gravity theories without projective symmetry \cite{BeltranJimenez:2019acz, BeltranJimenez:2020sqf}: the projective mode, which does not have a kinetic term, implies the existence of a ghost if all the fields propagate.

\subsubsection{Purely vector MAG with two propagating spin-1 fields} \label{sec:2vectors}

So far we have seen that it is not possible to have either four or three healthy propagating vectors. Let us explore the case of two. The starting point is, again, the matrix \eqref{eq:K3zeta0} and, in order to provide dynamical terms for two vectors, it must be positive semi-definite with just one zero eigenvalue. Due to this positive semi-definiteness, all the principal minors must be non-negative. If we compute the ones obtained after removing the second row and column, and the one coming from the elimination of the third ones, we get the conditions
\begin{equation}
    \det\left(\begin{array}{cc}0 & K_{1} \\K_{1} & K_{3}\end{array}\right)= -K_1^2 \geq 0 ,\qquad
    \det\left(\begin{array}{cc}0 & K_{2} \\K_{2} & K_{5}\end{array}\right)= -K_2^2 \geq 0 .
\end{equation}
This is only possible if $K_1$ and $K_2$ are vanishing, which implies the elimination of trace of the torsion from the kinetic sector. 

At this point we can collect all of our parameter restrictions:
\begin{align}
    \text{(IV)}&\coloneqq\text{(III)}\cup \{ \zeta \to 0,\ K_1\to 2,\ K_2\to 0 \}\nonumber\\
     &=\Big\{d_2\to d_5,\ d_3\to 4d_5,\ d_4\to 4d_5,\ d_6\to -\frac{1}{4}(d_1+2d_5),\nonumber\\
     &\qquad d_8\to -\frac{1}{2}(d_1+2d_5),\ d_9\to 2 d_7 ,\ d_{10} \to -\frac{1}{4}(d_1+2d_5),\nonumber\\
     &\qquad  d_{11} \to d_7,\ d_{13}\to d_{12} ,\  d_{15}\to 0,\ d_{16}\to \frac{1}{4} d_5\Big\}\,.
     \label{eq:condIV}
\end{align}
The resulting Lagrangian is:
\begin{eqnarray}
    \Lag_\text{v}|_\text{(IV)}&=& -\frac{1}{4}\ell^2 (\FD{Q}_{\mu\nu}, \FD{\vQ}_{\mu\nu}) \left(\begin{array}{cc} M_{3} & M_{4}\\ M_{4} & M_{5}\end{array}\right) \left(\begin{array}{c} \FD{Q}^{\mu\nu}\\ \FD{\vQ}^{\mu\nu}\end{array}\right)+\left(a_3-\frac{a_0}{6}\right)S_{\mu}S^{\mu}
    \nonumber\\
    &&+\left(a_2+\frac{2a_0}{3}\right)T_{\mu}T^{\mu}+\left(b_3+\frac{3a_0}{32}\right)Q_{\mu}Q^{\mu}+\left(b_4-\frac{a_0}{18}\right)\vQ_{\mu}\vQ^{\mu}
    \nonumber
    \\
    &&+\left(b_5-\frac{a_0}{4}\right)Q_\mu \vQ^\mu + \left(c_2-\frac{a_0}{2}\right)Q_\mu T^\mu +\left(c_3+\frac{2a_0}{3}\right)\vQ_\mu T^\mu \nonumber\\
    &&+\frac{4}{729}\ell^2\big(5d_1-4d_5+18d_7\big)(\vQ_\mu S^\mu)^2-\frac{8}{729}\ell^2\big(4d_1-5d_5+9d_7\big)\vQ_\mu\vQ^\mu S_\nu S^\nu \nonumber\\
    &&+\frac{2}{9}\ell^2 (d_5+d_7+d_{12}) (\star\FD{Q})_{\mu\nu}\vQ^\mu S^\nu+\frac{8}{81}\ell^2 (d_1-d_5+3d_7) (\star\FD{\vQ})_{\mu\nu}\vQ^\mu S^\nu,\label{eq:LagvafterIV}
\end{eqnarray}
where
\begin{align}
    M_3 &:= K_3|_\text{(IV)}=-\frac{1}{4}(d_1+d_5+d_7+2d_{12}+4d_{14}),\\
    M_4 &:= K_4|_\text{(IV)}=-\frac{1}{3}(d_5+d_7+d_{12}),\\
    M_5 &:= K_5|_\text{(IV)}=-\frac{4}{27}(d_1-d_5+3d_7).
\end{align}

After the previous parameter restrictions, the vectors $T_\mu$ and $S_\mu$ continue appearing in the vector Lagrangian $\Lag_{\text{v}}$ as auxiliary fields or Lagrange multipliers (depending on the values of the parameters). If we stick to the purely vector Lagrangian (i.e., we impose $\Lag_*=0$ in the full theory), one now has to integrate the corresponding equations of motion and/or use the resulting constraints in order to ensure that the remaining vectors ($Q_\mu$ and $\vQ_\mu$) propagate safely.

\subsection{Implications in the full theory}\label{sec:implicfullMAG}

Let us finally comment what we can learn from the previous analysis concerning the full theory, including the tensor sector $\Lag_*$. Firstly, notice that the conditions \eqref{eq:condIII} are still needed to ensure generically safe couplings between the different vectors and the graviton. Secondly, regarding the conditions obtained in Section \ref{sec:kinMAG}, one can also prove that these are also necessary conditions (according to our hypothesis of having at least one healthy propagating spin-1 field among $\{T_\mu,S_\mu,Q_\mu,\vQ_\mu\}$). It is true that we have only analyzed one part of the total kinetic matrix, since we have ignored the tensor parts and their kinetic couplings to vectors. However, the arguments we used to kill coefficients are based on the presence of non-positive numbers in the diagonal. A full kinetic sector that is positive definite (i.e., ghost-free) also requires the nullity of such coefficients. Therefore, we can conclude that the conditions  \eqref{eq:condIV} are necessary in the full theory. However, note that they are not sufficient, because the total stability requires also the analysis of $\Lag_*$. 

Finally, one interesting remark to mention is that, although $\{T_\mu,S_\mu\}$ appear as auxiliary fields or Lagrange multipliers in \eqref{eq:LagvafterIV}, this does not mean that they are non-dynamical in the full theory, because in principle they are also kinetically coupled to the tensor pieces. However, we do expect them to be non-dynamical, because the argument used in Section \ref{sec:2vectors} to eliminate $K_1$ and $K_2$ can also be applied to the other off-diagonal kinetic terms mixing $T_\mu$ and the tensor pieces. A detailed analysis of this question is left for future research.

\section{General quadratic Weyl-Cartan with propagating vector fields} \label{sec:Weyl}

Here we are going to perform a general analysis under the restrictions $\tQ_{\mu\nu\rho}=\vQ_\mu=0$. The resulting Lagrangian only depends on the three vector variables $T_\mu$, $S_\mu$ and $Q_\mu$, as well as on the tensor part $\tT_{\mu\nu}{}^\rho$. Under these conditions, the resulting Lagrangian, $\WC{\Lag}_\text{MAG}$, contains several invariants that are linearly dependent with the rest (up to boundary terms), so they can be dropped without loss of generality. To be precise, the following properties of Weyl-Cartan geometries\footnote{We are not putting a hat over the homothetic curvature because there is no difference between the general metric-affine one and the Weyl-Cartan one (see \eqref{eq:homothR}).}
\begin{equation}
    \WCR_{\mu\nu(\rho\lambda)}= \frac{1}{4}\homR_{\mu\nu}g_{\rho\lambda}\,,\qquad \WC{\coR}_{\mu\nu}=-\WCR_{\mu\nu}+\frac{1}{2}\homR_{\mu\nu}\,
\end{equation}
allow to eliminate the terms with $\{d_1,\,d_3,\,d_8,\,d_9,\,d_{10},\,d_{11},\,d_{13}\}$ from $\WC{\Lag}_\text{MAG}$. To remove the $d_4$ term, we notice that for these geometries
\begin{equation}
   4 R^{\mu\nu\rho\lambda} R_{\mu[\rho\lambda]\nu}= \WCR^{\mu\nu}\homR_{\mu\nu}-\frac{1}{4}\homR^{\mu\nu}\homR_{\mu\nu}-\WCR^{\mu\nu\rho\lambda}\WCR_{\mu\nu[\rho\lambda]}-\WCR^{\mu\nu\rho\lambda} \WCR_{\rho\lambda\mu\nu}-\frac{1}{4} \WCaR^2\,.
\end{equation}
Finally, the elimination of $d_5$ can be done thanks to the fact that the term $\WCR^{\mu\nu\rho\lambda}\WCR_{\rho\lambda\mu\nu}$ can be expressed in terms of others already present in the Lagrangian by using the Gauss-Bonnet invariant,
\begin{equation}
    \WC{\Lag}_\text{GB}= \WCR^2-4\WCR_{\mu\nu}\WCR^{\nu\mu}+\WCR_{\mu\nu\rho\lambda}\WCR^{\rho\lambda\mu\nu}\,,
\end{equation}
which is a boundary term in Weyl-Cartan geometry. In other words, the parameters
\begin{equation}
    \{d_1,\,d_3,\,d_4,\,d_5,\,d_8,\,d_9,\,d_{10},\,d_{11},\,d_{13}\}\label{eq:WCelimpar}
\end{equation}
can be absorbed in an appropriate redefinition of the rest.

After all of this, we can conclude that the most general Weyl-Cartan Lagrangian contained in \eqref{eq:LagMAG} is of the form:
\begin{eqnarray}
    2\kappa \Lag_\text{WC} &\coloneqq& -2\kappa\Lambda + a_0 \WCR + a_1 \tT_{\mu\nu\rho} \tT^{\mu\nu\rho} + a_2 T_\mu T^\mu + a_3 \aT_\mu \aT^\mu + b_3 Q_\mu Q^\mu + c_2 Q_\mu T^\mu \nonumber \\
&&  +\ell^2 \Big[ \WCd_2 \WCR^{\mu\nu\rho\lambda} \WCR_{\mu\nu[\rho\lambda]}  + \WCR^{\mu\nu} (\WCd_6 \WCR_{(\mu\nu)}+\WCd_7 \WCR_{[\mu\nu]}) \nonumber\\
&&\qquad + \homR^{\mu\nu} (\WCd_{12}\WCR_{\mu\nu}+\WCd_{14}\homR_{\mu\nu}) +\WCd_{15} \WCR^2+ \WCd_{16}\WCaR^2\Big]\,,
\label{eq:LagWC}
\end{eqnarray}
where the hatted parameters are the ones absorbing \eqref{eq:WCelimpar}.

As it was done in the general metric-affine case, it is convenient to introduce the analogous splittings
\begin{equation}
    \Lag_\text{WC} = \Lag_\text{GR} + \Lag_{\LCR\LCR} + \WC{\Lag}_\text{v} + \WC{\Lag}_*\,, \label{eq:LagWCSplit}
\end{equation}
and
\begin{equation}
     \WC{\Lag}_\text{v} = \WC{\Lag}_T + \WC{\Lag}_S + \WC{\Lag}_Q + \WC{\Lag}_\text{v, mix}\,, \label{eq:LagWCSplit2}
\end{equation}
where the different pieces of the splitting coincide with the ones in \eqref{eq:LagSplit} after evaluating in a Weyl-Cartan geometry.

First, in order to recover the Einstein-Hilbert Lagrangian in the Riemannian limit, we get the conditions
\begin{equation}
    \WC{\text{(I)}}\coloneqq \big\{\WCd_2 \to \WCd_{15}\,,\WCd_6 \to -4\WCd_{15}\big\}\label{eq:condIWC}
\end{equation}
(compare with \eqref{cond:3}).

\subsection{Pure vector sector of quadratic Weyl-Cartan}

We follow essentially the steps of the  Sections \ref{sec:MAGselfstab} and \ref{sec:MAGmixstab}. Then, from the stability of the separated vector sectors we get, as for the metric-affine case, that the parts $\WC{\Lag}_T|_{\WC{\text{(I)}}}$ and $\WC{\Lag}_Q|_{\WC{\text{(I)}}}$ are already of the Proca form, and that removing the dangerous terms of $\WC{\Lag}_S|_{\WC{\text{(I)}}}$ enforces
\begin{equation}
    \WCd_{15}=\WCd_{16}=0\,.\label{cond:WCaxial}
\end{equation}
Now one should be careful since, due to the absence of $\vQ_\mu$, there is no analogous of \eqref{cond:tr2Q}. Therefore, so far we have only required the substitutions
\begin{equation}
    \WC{\text{(II)}}\coloneqq \Big\{\WCd_2\to 0,\ \WCd_6\to 0,\ \WCd_{15}\to 0,\ \WCd_{16}\to 0\Big\}\,.
\end{equation}

Under these conditions the full vector Lagrangian reduces to a kinetic and a mass sector as follows:
\begin{eqnarray}
   2\kappa\WC{\Lag}_\text{v}|_{\WC{\text{(II)}}} &=& -\frac{1}{4}\ell^2\ (\WC{\mathcal{K}})_{AB}\FD{(A)}_{\mu\nu}\FD{(B)}^{\mu\nu} + \left(-\frac{a_0}{6}+a_3\right)S_\mu S^\mu + \left(\frac{2a_0}{3}+a_2\right)T_\mu T^\mu \nonumber\\
    &&+\left(\frac{3a_0}{32}+b_3\right)Q_\mu Q^\mu + \left(-\frac{a_0}{2}+c_2\right)Q_\mu T^\mu,
\end{eqnarray}
with $\FD{(A)}\coloneqq(\FD{S},\FD{T},\FD{Q})$ and where the kinetic matrix is given by
\begin{equation}
    \WC{\mathcal{K}}\coloneqq\left(\begin{array}{cccc}
\frac{1}{9}d_7 & 0 & 0 \\
0 & -\frac{4}{9}d_7 & \frac{1}{3}(d_7+d_{12})\\
0 & \frac{1}{3}(d_7+d_{12}) & -\frac{1}{4}(d_7+2d_{12}+4d_{14})
\end{array}\right).
\label{eq:WCKinMatrix}
\end{equation}

We observe the same problem that was present in Section \ref{sec:kinMAG}: the presence of the axial torsion introduces a ghost in the theory that can only be removed by taking $\WCd_7=0$. The resulting kinetic sub-matrix still does not allow the remaining vectors to propagate unless the non-diagonal term vanishes ($\WCd_{12}=0$), which eliminates the trace of the torsion as a dynamical field. We can sum up the restrictions so far in
\begin{equation}
    \WC{\text{(III)}}\coloneqq \Big\{\WCd_2\to 0,\ \WCd_6\to 0,\ \WCd_7\to 0,\ \WCd_{12}\to 0,\ \WCd_{15}\to 0,\ \WCd_{16}\to 0\Big\}\,.
\end{equation}

The result is the Lagrangian
\begin{eqnarray}
    2\kappa\WC{\Lag}_\text{v}|_{\WC{\text{(III)}}}&=&\frac{1}{4}\ell^2d_{14} \FD{Q}_{\mu\nu}\FD{Q}^{\mu\nu} + \left(a_3-\frac{a_0}{6}\right)S_\mu S^\mu + \left(a_2+\frac{2a_0}{3}\right)T_\mu T^\mu \nonumber\\
    &&+\left(b_3+\frac{3a_0}{32}\right)Q_\mu Q^\mu + \left(c_2-\frac{a_0}{2}\right)Q_\mu T^\mu\,,
\end{eqnarray}
in which we will require $d_{14}<0$ in order to prevent $Q_\mu$ from propagating ghostly modes.

\subsection{Implications in the full quadratic Weyl-Cartan}

The previous restrictions, $\WC{\text{(III)}}$, correspond to just keeping the term $\homR_{\mu\nu}\homR^{\mu\nu}$ in the original curvature-square sector of the Lagrangian \eqref{eq:LagWC}. From the definition \eqref{eq:homothR}, one easily realizes that this quadratic term belongs entirely to the vector sector. Consequently, the tensor part reduces to just a mass term,
\begin{equation}
    \WC{\Lag}_*|_{\WC{\text{(III)}}}= \frac{1}{2\kappa}\left(-\frac{a_0}{2}+a_1\right)\tT_{\mu\nu\rho}\tT^{\mu\nu\rho}\,,
\end{equation}
which implies
\begin{eqnarray}
  2\kappa\Lag_\text{WC}|_{\WC{\text{(III)}}}&=& 2\kappa\Lag_\text{GR} + 2\kappa\WC{\Lag}_\text{v}|_{\WC{\text{(III)}}}\\
  &=&-2\kappa\Lambda+a_0\LCR+\frac{1}{4}\ell^2d_{14} \FD{Q}_{\mu\nu}\FD{Q}^{\mu\nu} + \left(a_3-\frac{a_0}{6}\right)S_\mu S^\mu +\left(a_1-\frac{a_0}{2}\right)\tT_{\mu\nu\rho}\tT^{\mu\nu\rho} \nonumber\\
    &&+ \left(a_2+\frac{2a_0}{3}\right)T_\mu T^\mu +\left(b_3+\frac{3a_0}{32}\right)Q_\mu Q^\mu + \left(c_2-\frac{a_0}{2}\right)Q_\mu T^\mu \,.\label{eq:WCIIILag}
\end{eqnarray}
Let us finally discuss what happens with the non-dynamical fields $\{T_\mu, \tT_{\mu\nu}{}^\rho , S_\mu\}$. The axial vector $S_\mu$ is decoupled from the rest of the fields. When $6a_3=a_0$, it is absent, and if $6a_3\neq a_0$, $S_\mu$ becomes an auxiliary field whose equation of motion has $S_\mu=0$ as unique solution. The same happens for the tensorial part $\tT_{\mu\nu\rho}$ under the conditions $2a_1=a_0$ and $2a_1\neq a_0$, respectively. So both fields can be eliminated from the theory. Concerning the trace vector $T_\mu$, its coupling to the trace of the nonmetricity requires distinguishing three cases:
\begin{enumerate}
\item $2c_2=a_0$: similar situation as the one discussed above for $S_\mu$ and $\tT_{\mu\nu\rho}$. The vector $T_\mu$ simply drops from the theory and we get the Einstein-Proca theory
 \begin{equation}
     \Lag_1=-2\kappa\Lambda+a_0\LCR+\frac{1}{4}\ell^2d_{14} \FD{Q}_{\mu\nu}\FD{Q}^{\mu\nu} +\left(b_3+\frac{3a_0}{32}\right)Q_\mu Q^\mu . \label{eq:EProca1}
 \end{equation}
 \item $2c_2\neq a_0$ and $3a_2=-2a_0$: the torsion trace $T_\mu$ acts as a Lagrange multiplier that essentially sets the trace of the nonmetricity to zero. Consequently, we reach a degenerate case without vectors, i.e., GR is recovered.
 \item  $2c_2\neq a_0$ and $3a_2\neq - 2a_0$: then $T_\mu$ acts as an auxiliary field whose equation of motion gives:
 \begin{equation}
     T_\mu = \frac{3}{4}\frac{a_0-2c_2}{2a_0+3a_2} Q_\mu\,.
 \end{equation}
 If we plug this back in the Lagrangian \eqref{eq:WCIIILag}, we get the Einstein-Proca theory
 \begin{equation}
     \Lag_2=-2\kappa\Lambda+a_0\LCR+\frac{1}{4}\ell^2d_{14} \FD{Q}_{\mu\nu}\FD{Q}^{\mu\nu}
 + \left[-\frac{3}{16}\frac{(a_0-2c_2)^2}{2a_0+3a_2}  +\frac{3a_0}{32}+b_3\right]Q_\mu Q^\mu \,.\label{eq:EProca2}
 \end{equation}
\end{enumerate}
To sum up, we have seen that the tensor piece $\tT_{\mu\nu}{}^\rho$ of quadratic Weyl-Cartan is trivialized, as well as the axial torsion. Only $Q_\mu$ and $T_\mu$ survive, the latter being just either an auxiliary field or a Lagrange multiplier. In the non-trivial cases (i.e., different from GR) one recovers Proca Lagrangians for the Weyl vector $Q_\mu$. It is interesting to notice that the stability conditions are sufficiently strong to eliminate other healthy vector interactions apart from the standard kinetic and mass terms for $Q_\mu$.

\section{Final discussion} \label{sec:conclusions}

In this paper we have considered the quadratic metric-affine theory and explored the implications of imposing the absence of ghosts in the vector sector defined by $\{T_\mu,S_\mu,Q_\mu,\vQ_\mu\}$. We have first concentrated on the pure vector metric-affine gravity, in which the torsion is assumed to contain only the trace and the axial part, and the nonmetricity contains only its two traces. In order to demand the vector stability, we have followed the procedure specified at the beginning of Section \ref{sec:purevec}. After performing such a procedure, we have been able to reduce the parameter space of the quadratic curvature part, which is the one carrying the dynamics of the four vector variables, from 16 to just 5 parameters, namely $\{ d_1, d_7, d_{12}, d_{14}, d_{15}\}$. 

These same techniques can be used in the vector sector of the Weyl-Cartan restriction of the quadratic metric-affine action.  As a result of the procedure described in the previous paragraph, we found that all the torsion contributions (including the tensorial part) disappear from the theory and the only non-trivial (i.e., different from pure GR) results are the Einstein-Proca Lagrangians for $Q_\mu$ given in \eqref{eq:EProca1} (if $2c_2= a_0$) and \eqref{eq:EProca2} (if $2c_2\neq a_0$ and $3a_2\neq -2a_0$). The entire vector dynamics is controlled by the parameter $d_{14}$, which must be negative in order to ensure non-ghostly propagation. Moreover, the parameter $b_3$ can be adjusted in both cases to avoid tachyonic instabilities. The fact that such Proca theories are compatible with our stability requirements is actually a remarkable result (although expected by virtue of \eqref{eq:homothR}). Compare for example, with the Poincar\'e Gauge gravity case, in which the stability of vectors does not allow for any dynamics and fully trivializes both the vector and the tensor sectors (see Appendix \ref{app:PGcase} or the original publication \cite{BeltranMaldonado2020}).

At this point, we would like to enumerate some limitations of the present work:
\begin{enumerate}
    \item We have not considered odd parity invariants and one might ask whether they can help to solve the instability issues. Due to the evidence from the Poincar\'e Gauge gravity case in \cite{BeltranMaldonado2020}, there is no reason to think that this is possible.
    \item We have studied the quadratic Lagrangians constructed under the requirements described at the beginning of Section \ref{sec:MAGlag}, i.e., as postulated in the context of gauge gravity theories. This indeed justifies why do we get so restrictive conditions, since we are not considering all of the possible invariants up to dimension-4.
    \item In addition to the previous point, it is worth highlighting that our results are only valid in 4 dimensions. In higher dimensions, additional operators are expected to survive.
    \item Lastly, the method we use is oblivious to stable theories in which the vectors do not propagate. Examples of this are $R^2$ or $R^{(\mu\nu)}R_{\mu\nu}$, which can be seen as an $f(R)$ or a Ricci-Based theory, respectively.
\end{enumerate}

We highlight one more time the main message of Section \ref{sec:purevec}: although we have just focused on studying the vector sector of quadratic metric-affine gravity, the resulting conditions (see \eqref{eq:condIV}) are necessary for the stability of vector in the full theory, including the tensor couplings $\Lag_*$. We found that among the surviving parameters from the curvature square terms, namely $\{ d_1, d_7, d_{12}, d_{14}, d_{15}\}$, all except $d_{14}$ give contributions to the kinetic sector of the tensor parts $\tQ_{\mu\nu\rho}$ and $\tT_{\mu\nu}{}^\rho$ and their kinetic couplings to the vectors. The case $d_1=d_7=d_{15}=0$ leads to $d_{12}=0$, which is needed in order to prevent one of the vectors from being a ghost. As a result, one of them disappears and we arrive at a stable Einstein-Proca Lagrangian for $Q_\mu$. However, due to the size of the expressions for the tensor sector, we have not been able to find a rigorous proof in favor of $d_1=d_7=d_{12}=d_{15}=0$ as the only restriction that gives a stable theory. There could be further boundary terms to be extracted and one might arrive at a stable theory in which some of the tensor parts propagate.

Finally, it is worth remarking the fact that the tensor parts $\tQ_{\mu\nu\rho}$ and $\tT_{\mu\nu}{}^\rho$ (around Minkowski space) contain modes of spin greater than 1, in particular several spin-2. It is well-known that having more than one spin-2 field in a theory requires a very exhaustive tunning of the parameters to avoid instabilities and, in our case, these tensor parts are coupled in very involved ways to the rest of the fields of the theory. Since we only have the freedom to choose the four parameters $\{ d_1, d_7, d_{12}, d_{15}\}$, it does not look clear that we could be able to reach such a level of tunning. Moreover, the stability of just the vectors $\{T_\mu, S_\mu, Q_\mu,\vQ_\mu\}$ is enough to reduce enormously the space of parameters, as commented above. Therefore, we conjecture that having a fully stable theory  (with one of these vectors) is a strong enough requirement that would lead to $d_1=d_7=d_{12}=d_{15}=0$. Nonetheless, such a claim would have to be confirmed in future studies.

~
\begin{acknowledgments}
The authors would like to thank Jose Beltrán Jiménez and Adri\`a Delhom for useful discussions and feedback. This research was supported by the European Regional Development Fund through the Center of Excellence TK133 “The Dark Side of the Universe”. AJC was also supported by the Mobilitas Pluss post-doctoral grant MOBJD1035. FJMT is supported by the ``Fundaci\'on Ram\'on Areces''.
\end{acknowledgments}

\appendix

\section{Full expressions of the 1-vector restrictions}\label{app:completevectorLag}

Here we present the expression of the pure vector parts \eqref{eq:restrictVecL}, which, for convenience, have written in terms of $\{\alpha,\beta\}$ instead of $\{ d_2, d_6\}$:
\begin{eqnarray}
2\kappa\Lag_S
    &=&\frac{1}{36}\ell^2(-8d_5+2d_4-d_7+d_9-d_{11}-4\beta+4d_{15})\FD{S}_{\mu\nu}\FD{S}^{\mu\nu}\nonumber \\
    &&+\left(a_3-\frac{a_0}{6}\right)S_{\mu}S^{\mu}
      +\frac{1}{108}\ell^2(3\beta+\alpha) (S_{\mu}S^{\mu})^2
      +\frac{1}{9} \ell^2(d_4-4d_5+\alpha)\LC{G}_{\mu\nu}S^{\mu}S^{\nu}\nonumber \\
    &&+\frac{1}{24}\ell^2(24d_{16}-2d_5-d_4)\SD{S}^2
      -\frac{1}{18} \ell^2(4d_5-d_4+6\beta+\alpha) \LCR\, S_\mu S^\mu,\\
2\kappa\Lag_T
    &=&\frac{1}{9}\ell^2(4d_5-d_4+d_7-d_9+d_{11}+\alpha+4\beta-4d_{15})\FD{T}_{\mu\nu}\FD{T}^{\mu\nu}
       +\left(\frac{2a_0}{3}+a_2\right)T_{\mu}T^{\mu}\nonumber \\
    &&+\frac{2}{54}\ell^2(\alpha+3\beta)\left[ 4(T_{\mu}T^{\mu})^2 + 9\SD{T}^2+12 \LCR\, T_\mu T^\mu - 18 \SD{T}\LCR - 12\SD{T}T_\mu T^\mu\,\right]\,,\\
2\kappa \Lag_Q
    &=&\frac{1}{64}\ell^2(4d_1+4d_5+4d_7+8d_{12}+16d_{14}+\alpha+4\beta-4d_{15})\FD{Q}_{\mu\nu}\FD{Q}^{\mu\nu} +\left(\frac{3a_0}{32}+b_3\right)Q_{\mu}Q^{\mu}\nonumber \\
    &&+\frac{1}{1024}\ell^2(\alpha+3\beta) \left[3 (Q_{\mu}Q^{\mu})^2 +12\SD{Q}^2+64 \LCR\, Q_\mu Q^\mu +256\SD{Q}\LCR+24\SD{Q}Q_\mu Q^\mu\right]\,,\\
2\kappa\Lag_\vQ
    &=&\frac{1}{324}\ell^2(28d_1+36d_5-4d_3+8d_8+48d_{10}+36d_{11}+\alpha+4\beta-4d_{15})\FD{\vQ}_{\mu\nu}\FD{\vQ}^{\mu\nu}\nonumber \\
    &&+\left(b_4-\frac{a_0}{18}\right)\vQ_{\mu}\vQ^{\mu}\nonumber \\
    &&+\frac{1}{8748}\ell^2(144d_1+72d_3+132d_8+312d_{10}+37\alpha+59\beta-32d_{15}) (\vQ_{\mu}\vQ^{\mu})^2\nonumber \\
    &&+\frac{1}{108}\ell^2(4d_1+2d_3+2d_8+12d_{10}+7\alpha+19\beta+8d_{15})\SD{\vQ}^2\nonumber \\
    &&+\frac{1}{81} \ell^2(8d_1+4d_3+10d_8+12d_{10}+\alpha+3\beta-12d_{15})\LCR\, \vQ_\mu \vQ^\mu \nonumber \\
    &&+\frac{1}{9}\ell^2(d_8-2d_{10}-2\alpha-5\beta-4d_{15})\SD{\vQ}\LCR\nonumber \\
    &&+\frac{4}{81}\ell^2(4d_1+2d_3+3\alpha+14\beta+8d_8-14d_{15})\LC{G}_{\mu\nu}\vQ^\mu\vQ^\nu  \nonumber \\
    &&-\frac{1}{486}\ell^2(24d_1+12d_3+17d_8+62d_{10}-4\alpha-31\beta+4d_{15})\SD{\vQ}\vQ_\mu \vQ^\mu .
\end{eqnarray}

\section{Review of results in the metric-compatible case} \label{app:PGcase}

In this appendix we review the results about the stability of the metric-compatible limit of metric-affine gravity, Poincar\'e Gauge theory, for completeness. In \cite{Hayashi:1980qp} it was found that, apart from the usual graviton, the matter content of Poincar\'e Gauge gravity consists of two massive spin-2, two massive spin-1 and two spin-0 fields. In \cite{YoNester1,YoNester2}, the authors performed a Hamiltonian analysis of the theory and found that the only modes that could propagate without any pathological behavior were the two spin-0 with different parity. Recently, authors in \cite{BeltranMaldonado2020} arrived at the same conclusions using simpler arguments, based on the stability of the vector sector (as we do in the present text). In the following, we will explain the reasoning behind the latter.

First, we compute the metric-compatible limit of the MAG Lagrangian \eqref{eq:LagMAG},
\begin{eqnarray}
    2\kappa \Lag_\text{PG} &:=& 2\kappa \tor{\Lag}_\text{MAG} \nonumber\\
    &=& -2\kappa\Lambda + a_0 \tilde{R} + a_1 \tT_{\mu\nu\rho} \tT^{\mu\nu\rho} + a_2 T_\mu T^\mu + a_3 \aT_\mu \aT^\mu + \ell^2 d_5 \tor{\Lag}_\text{GB} \nonumber \\
&& +\ell^2 \Big[
\tord_2\torR^{\mu\nu\rho\lambda}\torR_{\mu\nu\rho\lambda} + \tord_6\torR^{\mu\nu}\torR_{(\mu\nu)}+\tord_7\torR^{\mu\nu}\torR_{[\mu\nu]}+\tord_{15}\torR^2+ \tord_{16}\tor{\aR}^2\Big] \,,
\label{eq:PGLag}
\end{eqnarray}
where we have absorbed the redundant parameters in $\{\tord_2, \tord_6, \tord_7, \tord_{15}, \tord_{16}\}$ and dropped the term with the torsionful Gauss-Bonnet invariant. 

Now we do the substitution $\tor{\text{(I)}}=\{\tord_2\to\tord_{15},\, \tord_6\to-4\tord_{15}\}$ (compare with \eqref{eq:condIWC}) to recover the Einstein-Hilbert limit. To impose the stability of the massive spin-1 fields in the particle spectrum, we again consider the pure vector sector containing the trace $T_\mu$ and the axial component $S_\mu$ of the torsion (as in \cite{BeltranMaldonado2020}), obtaining
\begin{eqnarray}
\tor{\Lag}_\text{v}|_{\tor{\text{(I)}}}&=&\frac{1}{9}\lambda\FD{T}_{\mu\nu}\FD{T}^{\mu\nu}-\frac{1}{36}(\lambda-\rho)\FD{S}_{\mu\nu}\FD{S}^{\mu\nu}+\frac12m_T^2T_\mu T^\mu+\frac12 m_S^2S_\mu S^\mu\nonumber\\
&&+\frac{2}{81}\rho S_\mu S^\mu T_\nu T^\nu+\frac{4}{81}\sigma(S_\mu T^\mu)^2 +\frac{\rho}{18}\big(2\LC{G}^{\mu\nu}S_\mu S_\nu+\LCR S_\mu S^\mu\big)\nonumber\\
&&-\frac{1}{27}\rho\big(\SD{T} S_\mu S^\mu - 2 \FD{S}_{\mu\nu} S^\mu T^\nu \big)-\frac{2}{27}\sigma\SD{S}S^{\mu}T_{\mu}+\frac{1}{72}(2\sigma+\rho)\SD{S}^2.
\label{eq:PGaction2}
\end{eqnarray}
Additionally, the involved parameters relate to the ones of the original Lagrangian as
\begin{align}
\lambda&:=\ell^2\tord_7\,, &m_T^2 &:=\frac{4}{3}a_0+2a_2\,,\nonumber\\
\rho&:=-4\ell^2\tord_{15} \,,        &m_S^2 &:=-\frac{a_0}{3}+2a_3\,,\nonumber\\
\sigma&:=\ell^2(-\tord_{15}+36\tord_{16})\,.
\end{align}

From the full vector Lagrangian \eqref{eq:PGaction2} written in this form, it is easy to identify the unstable terms: $\SD{S}^2$, whose elimination requires $2\sigma+\rho=0$, and the non-minimal coupling $\LCR S_\mu S^\mu$, which leads us to $\rho=0$. This in combination with the previous condition results in $\rho=\sigma=0$. There are actually other pathological terms, with the schematic form $\SD{W} V_\mu V^\mu$, but they disappear after the previous restrictions.

The stability conditions $\tor{\text{(II)}}:=\{\tord_2\to 0, \tord_6\to 0, \tord_{15}\to 0, \tord_{16}\to 0\}$ eliminate all the interactions and only leave the free quadratic part
\begin{equation}
   \tor{\Lag}_\text{v}|_{\tor{\text{(II)}}}=\frac{1}{9}\lambda\FD{T}_{\mu\nu}\FD{T}^{\mu\nu}-\frac{1}{36}\lambda\FD{S}_{\mu\nu}\FD{S}^{\mu\nu}+\frac{1}{2}m_T^2T_\mu T^\mu+\frac{1}{2} m_S^2 S_\mu S^\mu\,,
\end{equation}
where the kinetic terms for $T_\mu$ and $S_\mu$  have necessarily opposite signs, hence signaling the unavoidable presence of a ghost. Therefore, the only stable possibility is to exactly cancel both kinetic terms. Consequently, the entire vector sector becomes non-dynamical.

After imposing all of these conditions, which we summarize in $\tor{\text{(III)}}:=\{\tord_2\to 0, \tord_6\to 0, \tord_7\to 0, \tord_{15}\to 0, \tord_{16}\to 0\}$, the full quadratic Poincar\'e Lagrangian \eqref{eq:PGLag} becomes:
\begin{equation}
2\kappa\Lag_\text{PG}|_{\tor{\text{(III)}}}=-2\kappa\Lambda + a_0 \mathring{R} + \left(a_1-\frac{a_0}{2}\right) \tT_{\mu\nu\rho} \tT^{\mu\nu\rho} +\frac{1}{2}m_T^2 T_\mu T^\mu+\frac{1}{2} m_S^2 S_\mu S^\mu.
\end{equation}
After eliminating the torsional sector by using the corresponding equations of motion, we arrive at just GR. Again, the stability of the vector sector reduces the theory to the Einstein-Hilbert Lagrangian. Nevertheless, one can go beyond the procedure presented in this paper (see beginning of Section \ref{sec:purevec}) and look for stable theories in which the spin-1 fields in $\{T_\mu, S_\mu\}$ do not propagate. This means there is room, e.g. for the longitudinal modes of the vectors (a scalar and a pseudoscalar) to propagate safely, allowing for non-trivial torsion terms.
In fact, one can show that under the appropriate inequalities given in \cite{BeltranMaldonado2020}, the following subclass of the PG Lagrangian \eqref{eq:PGLag} is a stable bi-scalar theory where the longitudinal mode of the vectors propagate safely:
\begin{equation}
   \Lag_\text{PG,biscalar} =-2\kappa\Lambda+  a_0\torR+ a_1 \tT_{\mu\nu\rho} \tT^{\mu\nu\rho} + a_2 T_\mu T^\mu + a_3 \aT_\mu \aT^\mu+ \varpi_1 \torR^2+ \varpi_2\tor{\aR}^2\,.
\end{equation}
Finally, let us remark that non-linear extensions of quadratic PG gravity have been considered to stabilize some modes of the theory \cite{Aoki:2020rae}.

\bibliographystyle{JHEP}
\bibliography{references.bib}

\end{document}